\documentclass[11pt]{article}
\usepackage{moriond,epsfig}

\def\Journal#1#2#3#4{{#1} {\bf #2}, #3 (#4)}

\def\NIMA{{\em Nucl. Instrum. Methods} A}
\def\NPB{{\em Nucl. Phys.} B}

\def\PRL{\em Phys. Rev. Lett.}
\def\PRD{{\em Phys. Rev.} D}

\def\PRS{{\em Proc. Roy. Soc. Lond.} A}
\def\RMP{\em Rev. Mod. Phys.}
\def\EPJ{{\em Eur. Phys. J.} C}
\def\AIP{\em AIP Conf. Proc.}

\def\be{\begin{equation}}
\def\ee{\end{equation}}
\def\bea{\begin{eqnarray}}
\def\eea{\end{eqnarray}}

\begin{document}
\vspace*{4cm}
\title{RECENT HERMES RESULTS ON DVCS}

\author{ B. KRAUSS on behalf of the HERMES Collaboration }

\address{Universit{\"a}t Erlangen-N{\"u}rnberg, Physikalisches Institut II,
  Erwin-Rommel-Str. 1,\\
91058 Erlangen, Germany}

\maketitle\abstracts{
The interference of Deeply Virtual Compton Scattering (DVCS) and
Bremsstrahlung leads to a beam-charge asymmetry that can be observed for
exclusive photon production in the collision of high energy leptons and
nucleons/nuclei. Recent results for a hydrogen and a deuterium target are reported
and a consistent tendency for a rise of the cosine $\phi$
coefficient with momentum transfer $|t|$ has been found.}

\section{Introduction}

In high energy collisions of leptons and nucleons/nuclei there are two
mechanisms that contribute to the exclusive production of single photons at
leading order QED: Either the photon is emitted by the lepton, in which case
the process is called Bremsstrahlung or Bethe Heitler process~\cite{bh,mt}(BH), or it is emitted by the nucleon/nucleus or one of its components. 

\begin{figure}
\begin{center}
\psfig{figure=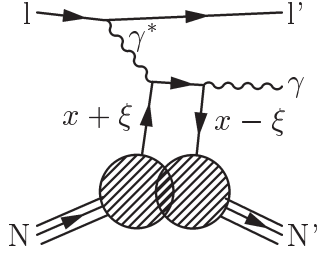,height=1.3in}
\end{center}
\caption{Deeply Virtual Compton Scattering: A virtual photon interacts with a
  quark from a nucleon/nucleus and the quark returns into the nucleon/nucleus after the
  emission of a real photon. 
\label{fig:fig0}}
\end{figure}

The description
of the latter process depends on the hadronic structure and it was shown~\cite{ji} that the relevant properties of
nucleons/nuclei can be parameterised by generalised parton
distributions (GPDs) under the condition that suitable scattering kinematics are selected: If the
four-momentum of the incoming (outgoing) lepton is denoted by $k$
($k^\prime$), the four-momentum of the incoming (outgoing) nucleon/nucleus is denoted
by $P$ ($P^\prime$) and the four-momentum of the real photon is denoted by $v$,
the following requirements have to be fulfilled: The four-momentum transfer
$Q^2=-(k^\prime - k)^2$ must be much larger than the mass
of the nucleon/nucleus, the final state mass $W^2=(v+P^\prime)^2$ must be outside the
resonance region and the four-momentum transfer $-t=-(q - v)^2$ must be
small compared with the target mass and $Q^2$.      
Under these conditions the latter process can be interpreted as Deeply Virtual
Compton Scattering (DVCS) (fig.\ref{fig:fig0}) and the absorption of the
virtual photon and the emission of the real photon are due to a
single quark. The probability of removing and reinserting the quark is parametrised by the GPDs.

In reality BH and DVCS mix due to quantum-mechanical interference and thus the
total cross-section involves a GPD-dependent interference term that is
sensitive to beam-charge and beam-polarisation for all
unpolarised hadronic targets. 

For the nucleon 4 leading twist GPDs exist, namely $\mathrm{H}$,
$\mathrm{E}$, $\mathrm{\tilde H}$ and $\mathrm{\tilde E}$. Each of them
depends on 3 variables: the lightcone momentum fraction $x$ of the quark
inside the nucleon/nucleus, the lightcone momentum transfer $\xi$ and
the value of $t$. GPDs are constrained by
conventional quark distributions and nucleon formfactors and can
also be used to describe hard exclusive meson production.

For other targets, different numbers of GPDs are required, e.g. one GPD for a
spin-0 target like neon, or nine GPDs for a spin-1 target like deuterium. In
addition for all nuclear targets BH/DVCS can also take place on single
nucleons, especially at larger momentum transfers $|t|$. Apart from binding
effects in this case one effectively probes the GPDs of the nucleon. Moreover, this incoherent process can also occur with simultaneous
excitation of the nucleon into a resonant state. Also the coherent formation
of nuclear excitations may be possible. If the hadronic final state is
not explicitly detected, all these processes are difficult to discriminate experimentally.

\section{Experimental Setup and Previous Results}

HERMES \cite{he} is a fixed-target experiment based on a
multi-purpose forward spectrometer. Its internal gas target has been operated with various
unpolarised gases (e.g. H, D, Ne) as well as vector-polarised hydrogen and
 vector and tensor-polarised deuterium.        

In the case of BH/DVCS typically only the lepton and the photon are inside the
acceptance of the spectrometer. While the lepton is detected and identified by
several driftchambers as well as a transition radiation detector (TRD), a
preshower hodoscope (H2) and the electromagnetic
calorimeter, the photon is only detected by H2 and the calorimeter. The
requirement is imposed that no other charged tracks or untracked clusters in
the calorimeter are present in the same event. In addition standard fiducial
volume cuts are applied and the energy transfer
$\nu$ to the virtual photon in the laboratory system is required to be less
than 22~GeV for trigger stability reasons. Furthermore cuts on the event
kinematics are applied: As discussed above,
$Q^2>1$~GeV$^2$, $W^2>9$~GeV$^2$ and $-t<0.7$~GeV$^2$ are required. For
acceptance reasons the angle $\theta_{\gamma^* \gamma}$ between the real and
the virtual photon (assuming DVCS) is restricted to be less than
$45$~mrad, while detector smearing is considered to be acceptable above
$\theta_{\gamma^* \gamma}= 5$~mrad. A cut on the photon energy of
$E_\gamma>3$~GeV suppresses photons from meson decays.

In order to ensure exclusivity of an event the missing mass $M_x$ of
the hadronic final state is calculated under the assumption of a target
proton that was at rest. If $M_x^2$ is found to be negative, $M_x$
is defined as $-\sqrt{-M_x^2}$. For exclusive events $-1.5$~GeV$<M_x<+1.7$~GeV is required, which selects coherent as
well as incoherent BH/DVCS events for all nuclear targets. The
background of other event types - mainly photons from $\pi^0$-decay - is
determined from a data/Monte Carlo comparison and subtracted
after estimating the background asymmetry.

The remaining event sample for a Deuterium target is dominated by incoherent, elastic
BH/DVCS on the proton, which contributes approximately 50\% to the event statistics for all
values of $t$. Coherent BH/DVCS on the deuteron only has a sizable contribution
to the event sample in the lowest $|t|$-region (about 40\% in the lowest
$|t|$-bin). The remaining fraction is shared between the
incoherent process with resonance excitation and the incoherent, elastic
process on the neutron.

Already some time ago HERMES has reported the observation of a significant
single beam-spin asymmetry on Deuterium as well as on other targets
using similar data selection criteria \cite{ha,hb}. These results were supported by
measurements at JLAB \cite{cl} which showed a beam-spin asymmetry of
comparable size but the
opposite sign, since the opposite beam-charge is used.  

\section{Beam-charge Asymmetry on Hydrogen and Deuterium}
The beam-charge asymmetry $A_C$ of BH/DVCS may be even more interesting for the following reason: It
is known that the (here) dominant GPDs $H^q$ approach the ordinary quark
distributions as function of $x$ in the limit of $\xi \rightarrow 0$ and $t
\rightarrow 0$. The negative $x$ region corresponds to the antiquark
distribution as a function of $|x|$ and with negative sign. 

Far away from this kinematic limit only the integral over $x$ is
constrained by the Dirac formfactor of the nucleon. This condition does not
restrict contributions to $H^q$ that are entirely odd in $x$ and there are at
least two candidates for this: The contribution from sea-quarks and the
so-called D-term \cite{pw}. Little is known about them and
$A_c$ picks out exactly these parts.

For the following results the beam-charge asymmetry on deuterium has been
calculated as 
\begin{equation}
A_C(\phi)=\frac{N^+(\phi) - N^-(\phi)}{N^+(\phi) + N^-(\phi)}
\end{equation}
in 10 bins in $\phi$, where $\phi$ denotes the azimuthal
orientation of the real photon about the direction of the virtual photon
and $N^+$ and $N^-$ are the luminosity-weighted
event numbers for positive and negative beam-charge, respectively. 
The asymmetry has been fitted by the following function:
\begin{equation}
A(\phi) = A_C^{const.} + A_C^{\cos \phi} \cos(\phi) + A_C^{\sin \phi} \sin(\phi).
\end{equation}

\begin{figure}
\begin{center}
\psfig{figure=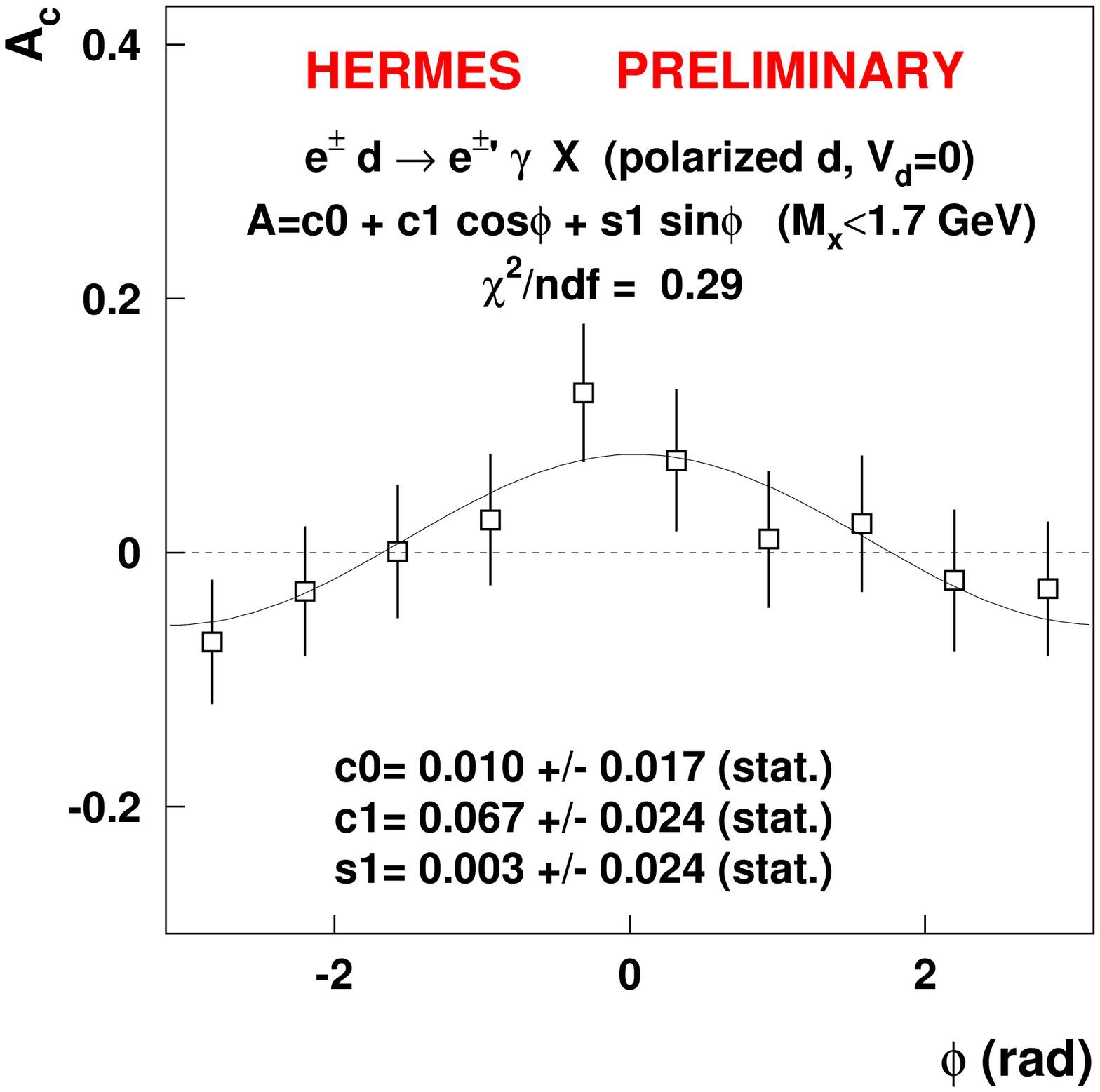,height=2.in}
\psfig{figure=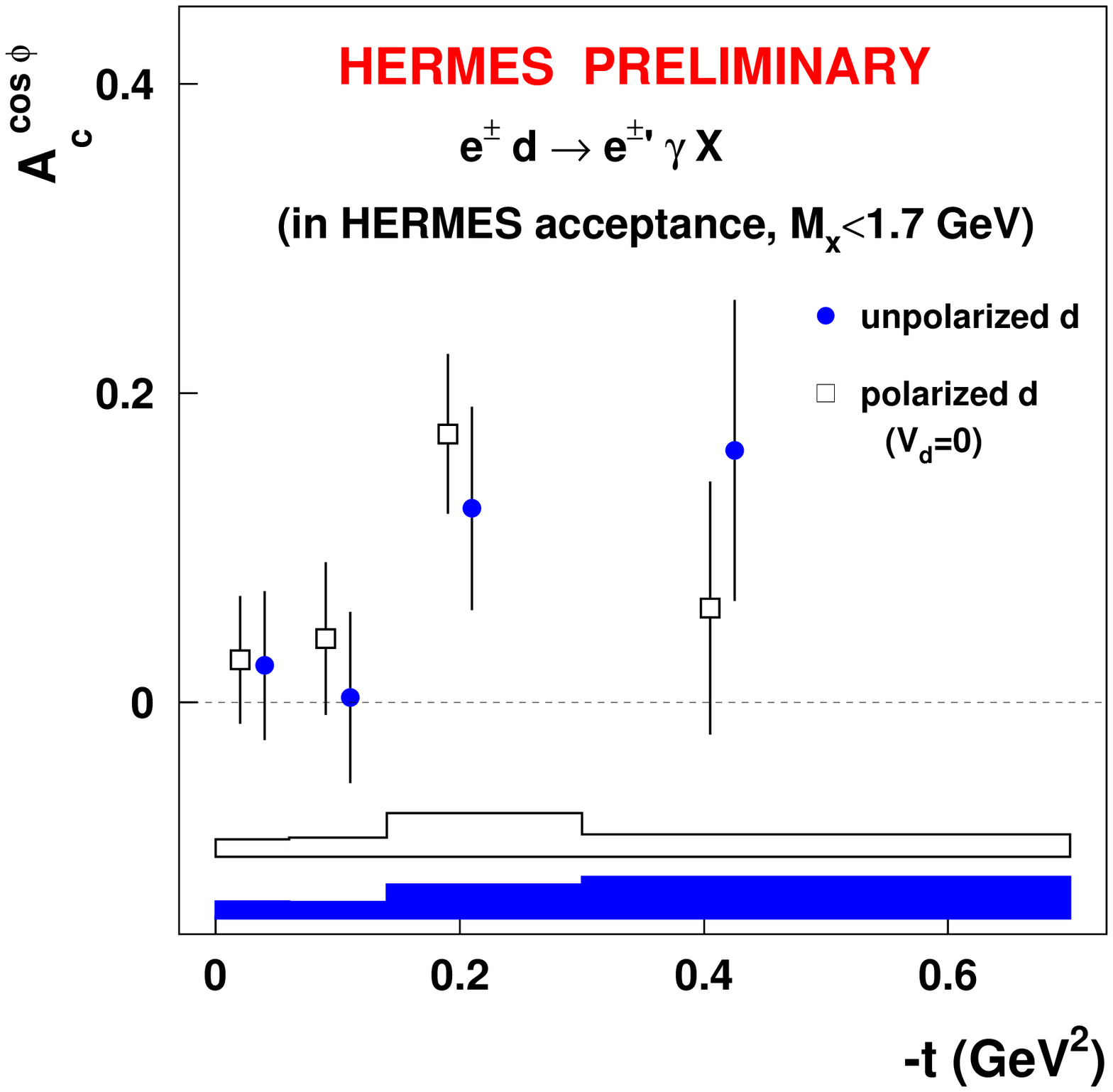,height=2.in}
\end{center}
\caption{The beam-charge asymmetry on tensor-polarised deuterium as a function
  of the angle $\phi$ (left) and the $\cos(\phi)$ coefficient of the
  beam-charge asymmetry on unpolarised/tensor-polarised deuterium as a
  function of $|t|$(right).
\label{fig:fig1}}
\end{figure}

Two different deuterium datasets exist at HERMES: one dataset is
taken with an unpolarised target and one dataset has a vanishing vector
polarisation $V_d$, but a large positive tensor polarisation. Figure
\ref{fig:fig1} (left)
shows the beam-charge asymmetries from the second subset as a function of $\phi$; only statistical errors are plotted. For the
coefficient $A_C^{\cos \phi}$ a $2.5~\sigma$ deviation from zero is found. 

As this asymmetry arises from a mixed dataset, it is more
instructive to study the asymmetry as a function of $t$. This is shown in
figure \ref{fig:fig1} (right). For both datasets there is a tendency that the asymmetry
rises to positive values for large values of $|t|$. In the low $|t|$-range
there could be a difference due to tensor-effects in the coherent process, but
such a difference is not observed. It has to be noted that due to the
limited statistics and the comparatively large kinematical bins in $t$ there is an
additional model-dependent binning error on the coefficient $A_C^{\cos \phi}$
that is estimated to be about 20 to 30 per cent in comparison with the true
asymmetry at bin center. 

\begin{figure}
\begin{center}
\psfig{figure=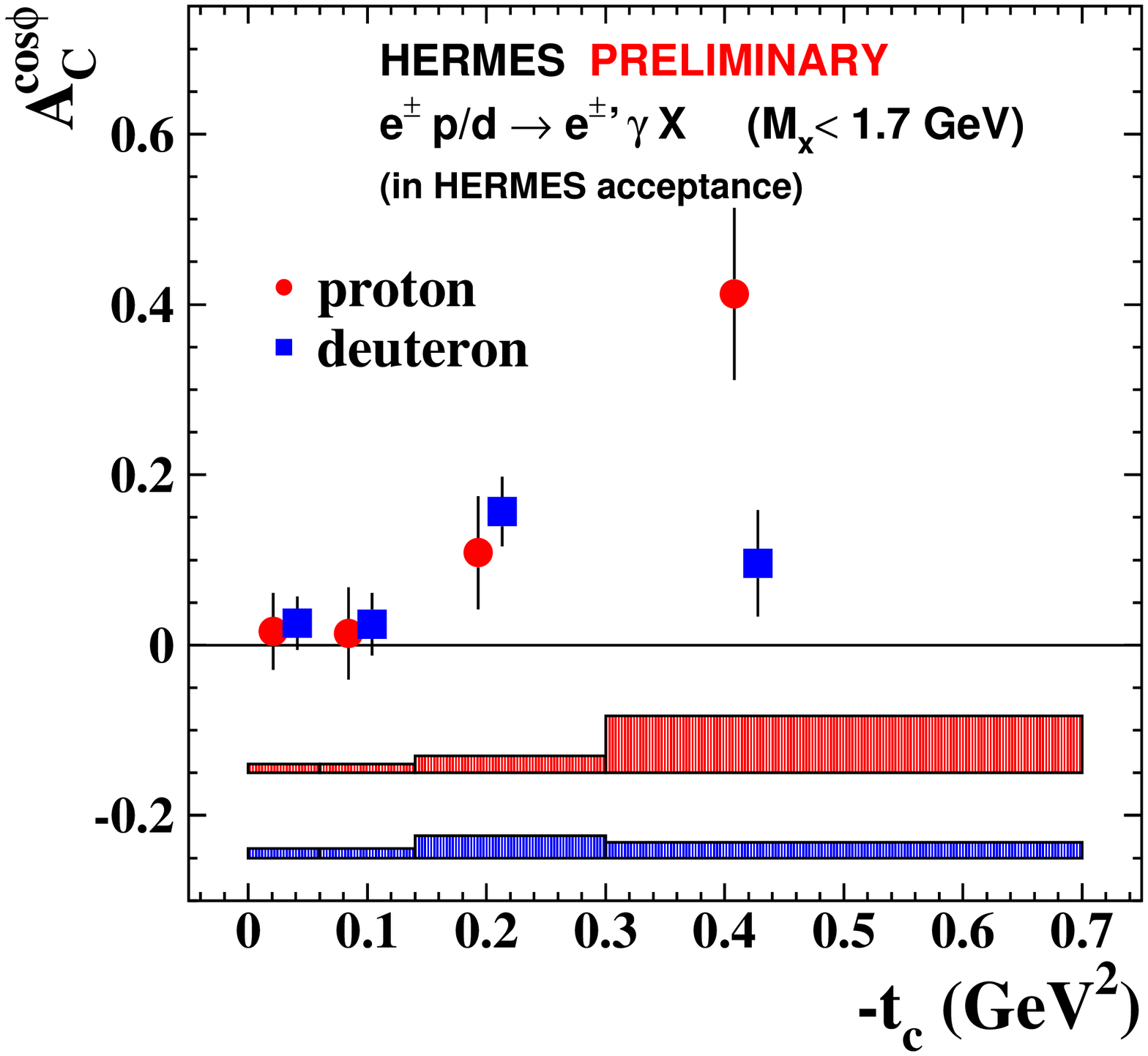,height=2.0in}
\psfig{figure=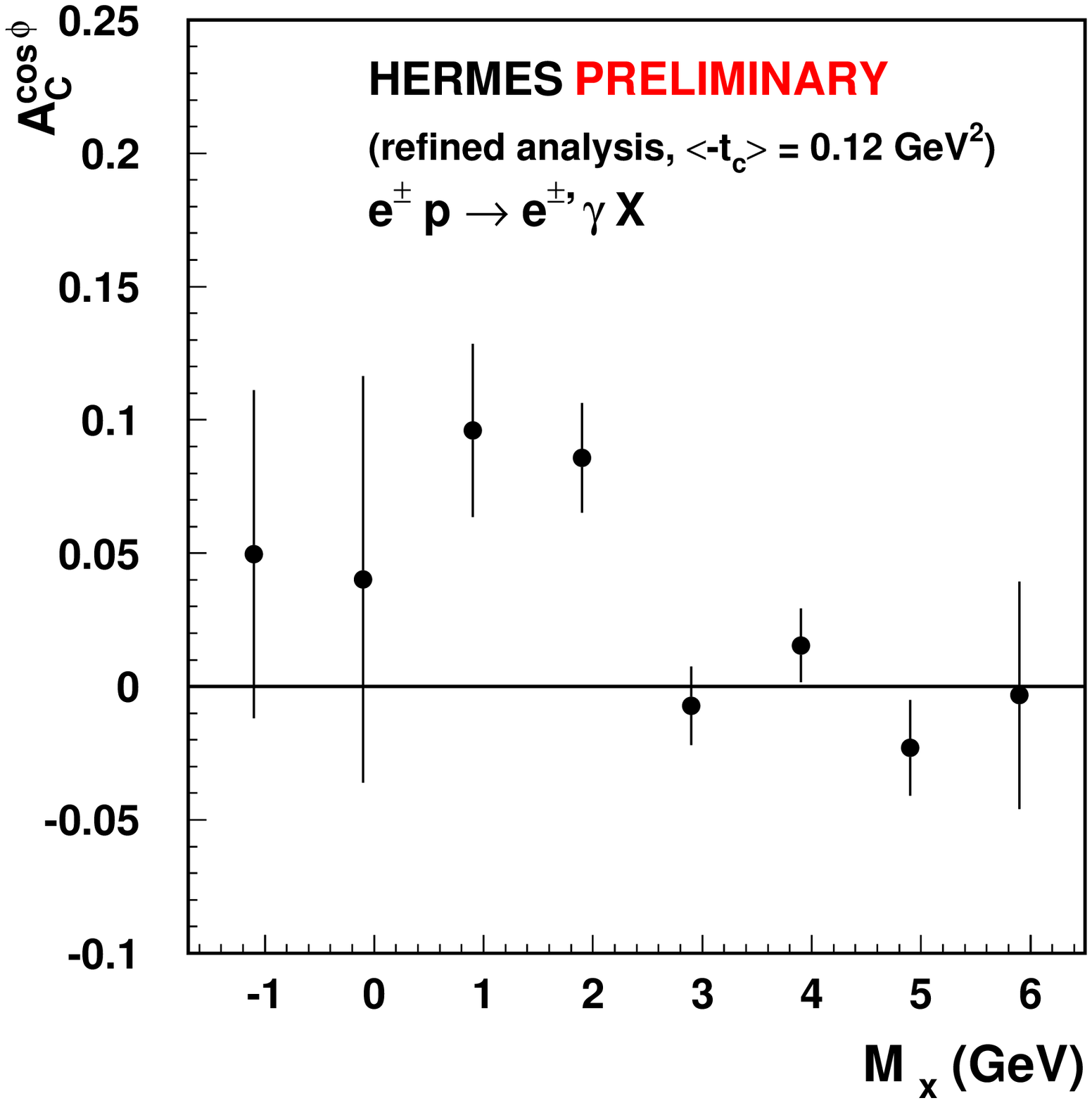,height=2.0in}
\psfig{figure=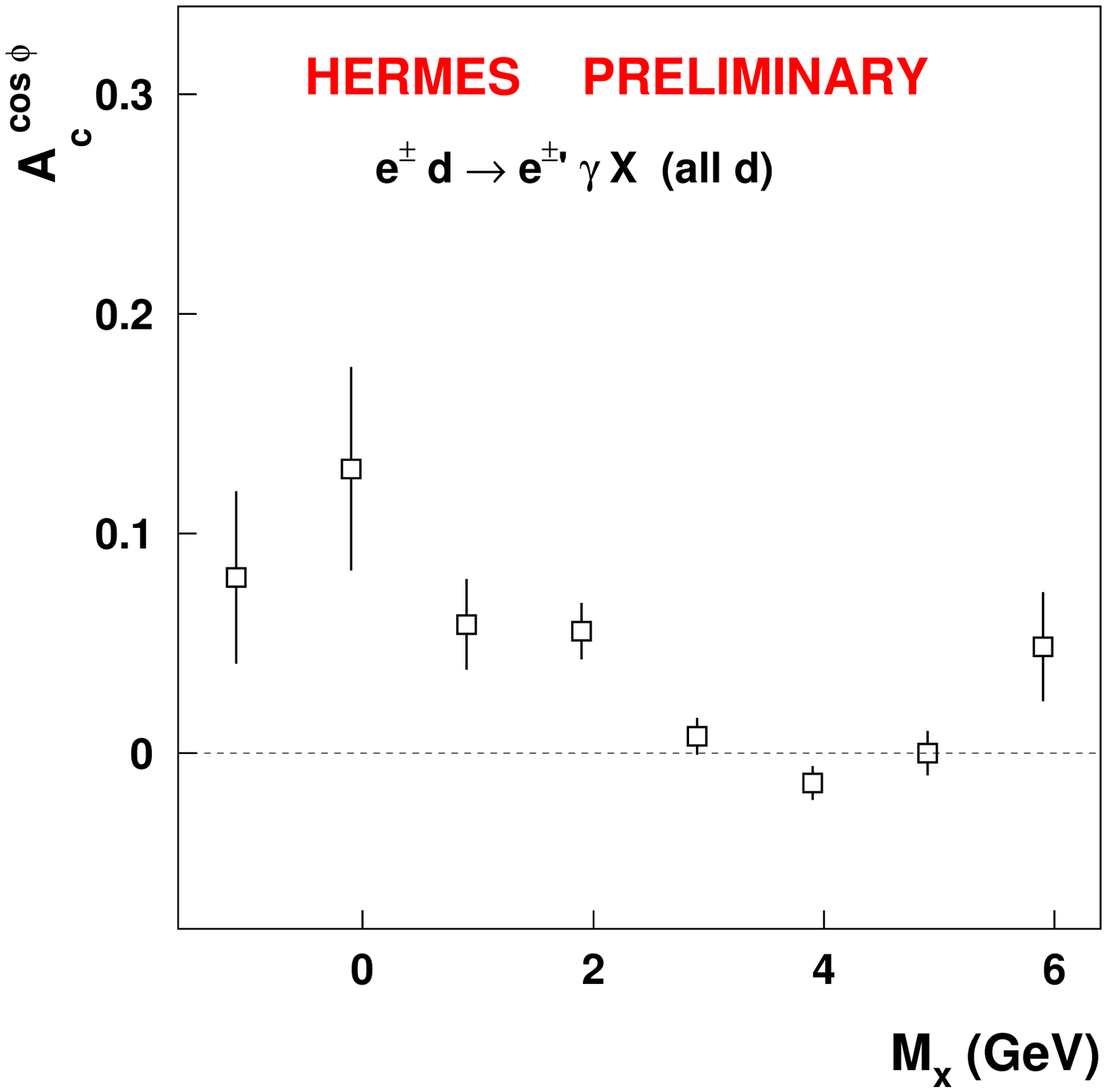,height=2.0in}
\end{center}
\caption{$t$-dependence of the $\cos(\phi)$ coefficient of the beam-charge
  asymmetry for unpolarised hydrogen in comparison with unpolarised/tensor-polarised deuterium (left). Cosine coefficient of the beam-charge
  asymmetry on hydrogen (middle) and unpolarised/tensor-polarised deuterium (right) as a function of the missing
  mass $M_x$.
\label{fig:fig3}}
\end{figure}

As shown in figure \ref{fig:fig3} (left) the obtained results are 
consistent with the proton results that have been extracted using the same
procedure. The apparent difference in the highest bin in $|t|$ can be due to
the resonance contributions, which may be different for the proton and the
neutron but are essentially unknown.

As an additional cross-check the coefficient $A_C^{\cos \phi}$ has been obtained as
a function of the missing mass. Figure \ref{fig:fig3} (middle, right) demonstrates
that the asymmetry is only found to be non-zero for the exclusive missing mass range - as expected. 

The observed results for the beam-charge asymmetry on deuterium can
approximately be described by Monte Carlo models \cite{kn}.
However, no sea-quark contribution is at present included in these models and
also the DVCS amplitude for resonance excitation is omitted. Thus there exists considerable freedom to adjust the model predictions to the
observed results.

\end{document}